\documentclass[11pt,twoside]{article}
\usepackage{./asp2014}

\resetcounters

\begin{document}

\title{Resolved Substructures in Protoplanetary Disks with the ngVLA}

\author{Sean M.~Andrews,$^{1}$ David J.~Wilner,$^{1}$ Enrique Mac{\'\i}as,$^{2}$ Carlos Carrasco-Gonz{\'a}lez,$^{3}$ and Andrea Isella$^{4}$ }
%\author{Sean M.~Andrews, David J.~Wilner,}
\affil{$^1$Harvard-Smithsonian Center for Astrophysics, Cambridge, MA, USA 02138; \email{sandrews@cfa.harvard.edu},  \email{dwilner@cfa.harvard.edu}}
%\author{Enrique Mac{\'\i}as,}
\affil{$^2$Department of Astronomy, Boston University, Boston, MA, USA 02215; \email{emacias@bu.edu}}
%\author{Carlos Carrasco-Gonz{\'a}lez,}
\affil{$^3$Instituto de Radioastronom{\'\i}a y Astrof{\'\i}sica (IRyA-UNAM), 58089, Morelia, M{\'e}xico; \email{c.carrasco@irya.unam.mx}}
%\author{and Andrea Isella}
\affil{$^4$Department of Physics and Astronomy, Rice University, Houston, TX, USA 77005; \email{isella@rice.edu}}

%\author{Sean M.~Andrews, David J.~Wilner,}
%\affil{Harvard-Smithsonian Center for Astrophysics, Cambridge, MA, USA 02138; \email{sandrews@cfa.harvard.edu},  \email{dwilner@cfa.harvard.edu}}
%\author{Enrique Mac{\'\i}as,}
%\affil{Department of Astronomy, Boston University, Boston, MA, USA 02215;
%\email{emacias@bu.edu}}
%\author{Carlos Carrasco-Gonz{\'a}lez,}
%\affil{Instituto de Radioastronom{\'\i}a y Astrof{\'\i}sica (IRyA-UNAM), 58089, Morelia, M{\'e}xico;
%\email{c.carrasco@irya.unam.mx}}
%\author{and Andrea Isella}
%\affil{Department of Physics and Astronomy, Rice University, Houston, TX, USA 77005; \email{isella@rice.edu}}

\paperauthor{Sean M. Andrews}{sandrews@cfa.harvard.edu}{0000-0003-2253-2270}{Harvard-Smithsonian Center for Astrophysics}{}{Cambridge}{MA}{02138}{USA}
\paperauthor{David J. Wilner}{dwilner@cfa.harvard.edu}{0000-0003-1526-7587}{Harvard-Smithsonian Center for Astrophysics}{}{Cambridge}{MA}{02138}{USA}
\paperauthor{Enrique Macias}{emacias@bu.edu}{0000-0003-1283-6262}{Deparment of Astronomy, Boston University}{}{Boston}{MA}{02215}{USA}
\paperauthor{Carlos Carrasco-Gonzalez}{c.carrasco@irya.unam.mx}{}{Instituto de Radioastronomia y Astrofisica}{}{Morelia}{}{58089}{Mexico}
\paperauthor{Andrea Isella}{isella@rice.edu}{0000-0001-8061-2207}{Department of Physics and Astronomy, Rice Univeristy}{}{Houston}{TX}{77005}{USA}

%%EJM -- received from Sean on 8/16/18
\begin{abstract}
Terrestrial planets and the cores of giant planets are thought to be built by the collisional agglomeration of solids spanning over 20 orders of magnitude in size within a few million years.  However, there is tension between this basic picture of planet formation and standard theoretical assumptions associated with the migration of ``pebbles" ($\sim$mm/cm-sized particles) in gas-rich disks and the presumably much longer timescales necessary to assemble ($\sim$km-scale) ``planetesimals".  To confront these potential theoretical discrepancies with observational constraints, the ideal tracer of the solids concentrated in protoplanetary disk substructures is the $30-100$\,GHz continuum, which strikes the best balance in sensitivity (emission still bright), optical depth (low enough to reliably estimate densities), and angular resolution (high enough to resolve fine-scale features at disk radii as small as 1 au).  With its combination of sensitivity, frequency coverage, and angular resolution, the next-generation VLA will be the only facility that has the capabilities to open up this new window into physics of planetesimal formation.
\end{abstract}

\section{Introduction}

In the ``core accretion" paradigm for planet formation, terrestrial planets and giant planet cores are created by the sequential collisional agglomeration of solids over $\sim$20 orders of magnitude in size within a few million years \citep[e.g.,][]{pollack96,raymond14}.  Early in that process, the standard theoretical assumptions introduce two fundamental obstacles to that growth.  The first is related to the migration of ``pebbles" ($\sim$mm/cm-sized particles): as these solids decouple from the gas disk, they migrate faster than they can collide and grow \citep{takeuchi02,brauer07}.  The second is that the timescales for assembling ($\sim$km-scale) ``planetesimals" is too long, given the migration and destructive impacts of their precursors \citep[e.g.,][]{johansen14}.  The potential solution to both issues is to locally concentrate pebbles in the disk \citep{whipple72,pinilla12a}, halting their migration and slowing their relative velocities to promote rapid growth \citep[e.g.,][]{chiang10}. The requirement to facilitate such concentrations is a gas pressure profile that is not smooth or monotonic, but rather has local maxima induced by abrupt variations in disk properties \citep[e.g.,][]{dzyurkevich10,stammler17}, dynamical effects \citep[e.g.,][]{baruteau14}, or fluid instabilities \citep[e.g.,][]{zhu14,flock15}.

The optimal way to directly test this hypothesis is to search for and characterize local concentrations of pebbles in protoplanetary disks.  These particles emit a thermal continuum at wavelengths roughly comparable to their sizes (i.e., radio to microwave frequencies, $\sim$10--1000\,GHz); the detailed spatial distribution of this continuum emission in nearby disks can be resolved with an interferometer.  If the optical depth of this emission is low, the spectrum provides constraints on the temperatures, densities, and size distribution of the constituent solids in any accessible disk region.  Variations in the particle concentrations on small spatial scales would be manifested as {\it substructures} in the microwave continuum surface brightness morphology.  The properties of those substructures provide crucial information on the mechanisms that create them and their potential for driving rapid, localized planetesimal formation.  

\begin{figure*}[!t]
\includegraphics[width=\linewidth]{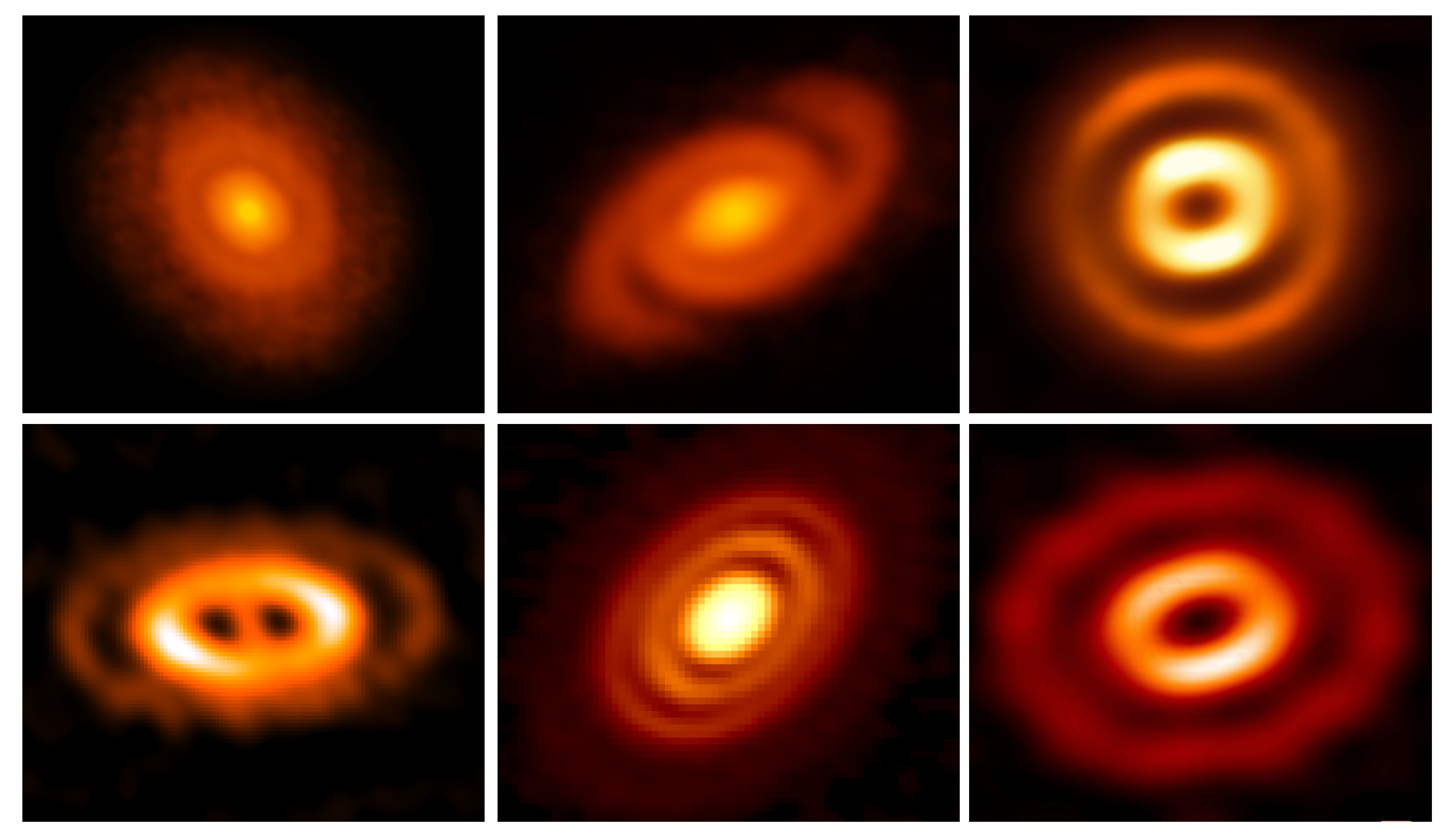}
\caption{Examples of substructures in the $\sim$1\,mm continuum emission from nearby protoplanetary disks at scales of $\sim$10--20\,au, from the ALMA interferometer.  Clockwise from top left: the V883 Ori \citep{cieza16}, Elias 27 \citep{perez16}, HD 169142 \citep{fedele17}, HD 97804 \citep{vanderplas17}, HD 163296 \citep{isella16}, and AA Tau \citep{loomis17} disks.}
\end{figure*}

In the past few years, the ALMA interferometer has found that such substructures on scales of $\lesssim$10--20\,au (70--150\,mas) are common, and perhaps ubiquitous \citep[e.g.,][]{zhang16}.  Usually they are manifested as concentric bands or narrow rings of emission separated by pronounced depletions, or ``gaps" \citep{isella16,cieza16,cieza17,loomis17,vanderplas17,cox17,fedele17,fedele18,dipierro18}, although at least one disk has a spiral emission pattern \citep{perez16}.  Figure 1 shows a gallery of representative ALMA continuum images of disk substructures.  In the two specific cases of the HL Tau and TW Hya disks, ALMA data reveal ring+gap features at even finer scales, down to (and presumably beyond) the $\sim$20\,mas (3 and 1\,au, respectively) resolution limit of the longest ALMA baseline configuration \citep{brogan15,andrews16}.

\section{The Role of the ngVLA}

\begin{figure*}[!t]
\includegraphics[width=\linewidth]{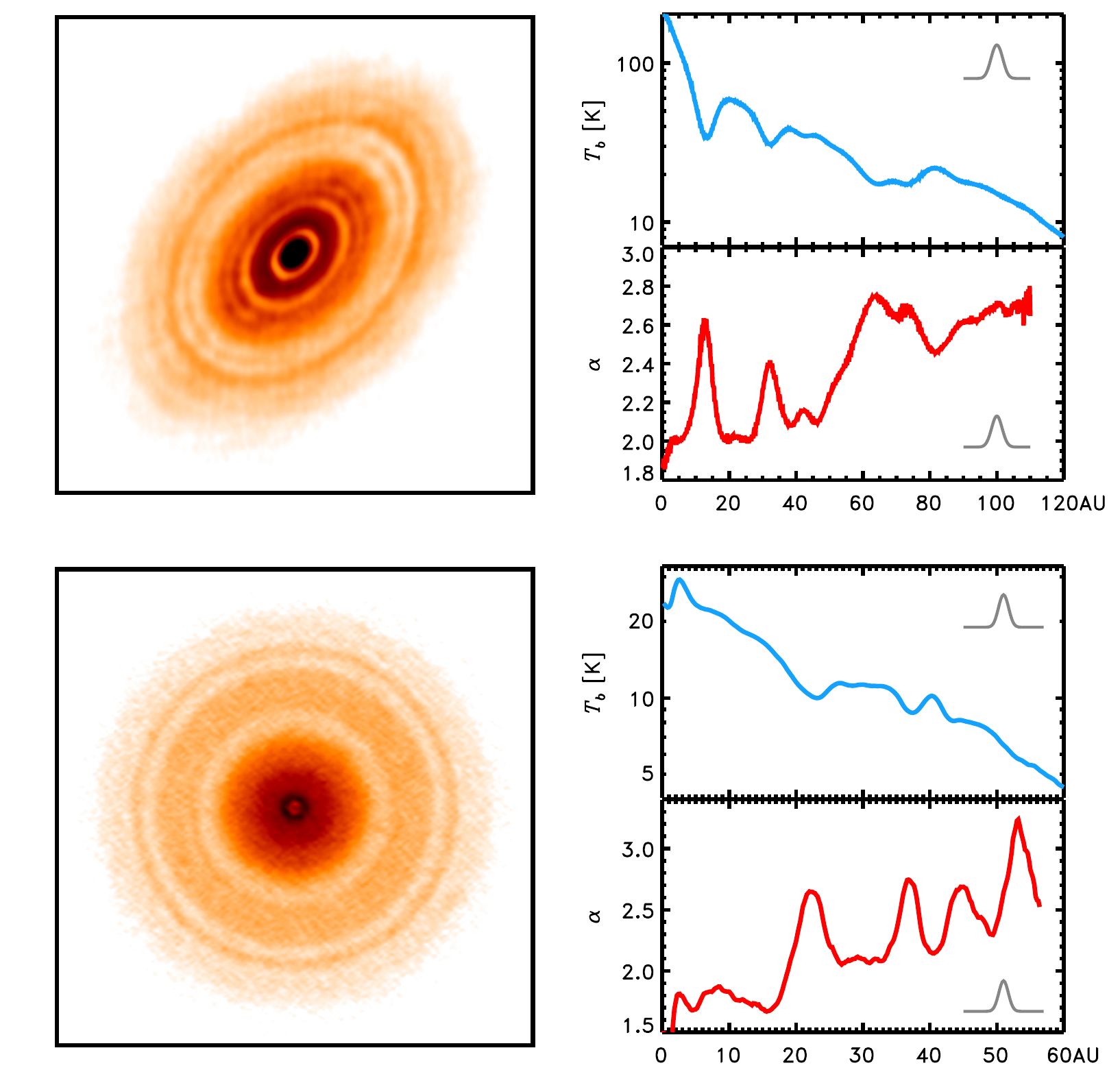}
\caption{A more detailed look at ALMA observations of the HL Tau ({\it top}) and TW Hya ({\it bottom}) disks, based on the analyses of \citet{brogan15} and \citet{huang18}.  The left panels show 290\,GHz continuum images.  The right panels feature the azimuthally-average radial $T_b$ ({\it top}) and $\alpha$ (340--230\,GHz) profiles.  The corresponding beam profiles are shown in gray.  Local $T_b$ depletions are accompanied by $\alpha$ enhancements.  The high $T_b$ values and low $\alpha$ values in the emission bands / rings are suggestive of high optical depths at these frequencies.}
\end{figure*}

While these ALMA discoveries have rightly ushered in a major shift in the field, they also reveal two crucial, but subtle, problems related to the interpretation of disk substructures.  First, the emission substructures that have so far been observed at very high resolution appear to be largely optically thick at frequencies higher than $\sim$200\,GHz.  And second, the ring or gap features often remain unresolved (or nearly so) even for the longest ALMA baselines (scales $\lesssim$20\,mas).  Both of these issues are illustrated in Figure 2, which shows synthesized images along with azimuthally-averaged brightness temperature ($T_b$) and spectral index ($\alpha$) profiles for the HL Tau ({\it top}) and TW Hya ({\it bottom}) disks for the continuum emission around 290\,GHz (\citealt{brogan15}; \citealt{huang18}; see also \citealt{tsukagoshi16}).  Note that the scales of the substructures are comparable to the resolutions in the $T_b$ profiles, and also that the $T_b$ values are similar to the expected dust temperatures at the locations where the spectral indices are near the Rayleigh-Jeans value, as would be expected for high optical depths ($\alpha \approx 2$).  If this behavior ends up being common in the broader disk population, then high frequency continuum measurements will only be able to offer relatively weak limits on the densities and particle size distributions in these substructures.  Moreover, all but the strongest features in the innermost disk ($\lesssim$10\,au) will be inaccessible at ALMA frequencies, due to both very high optical depths and limited angular resolution.    

\begin{figure*}[!t]
\begin{center}
    \includegraphics[width=4in]{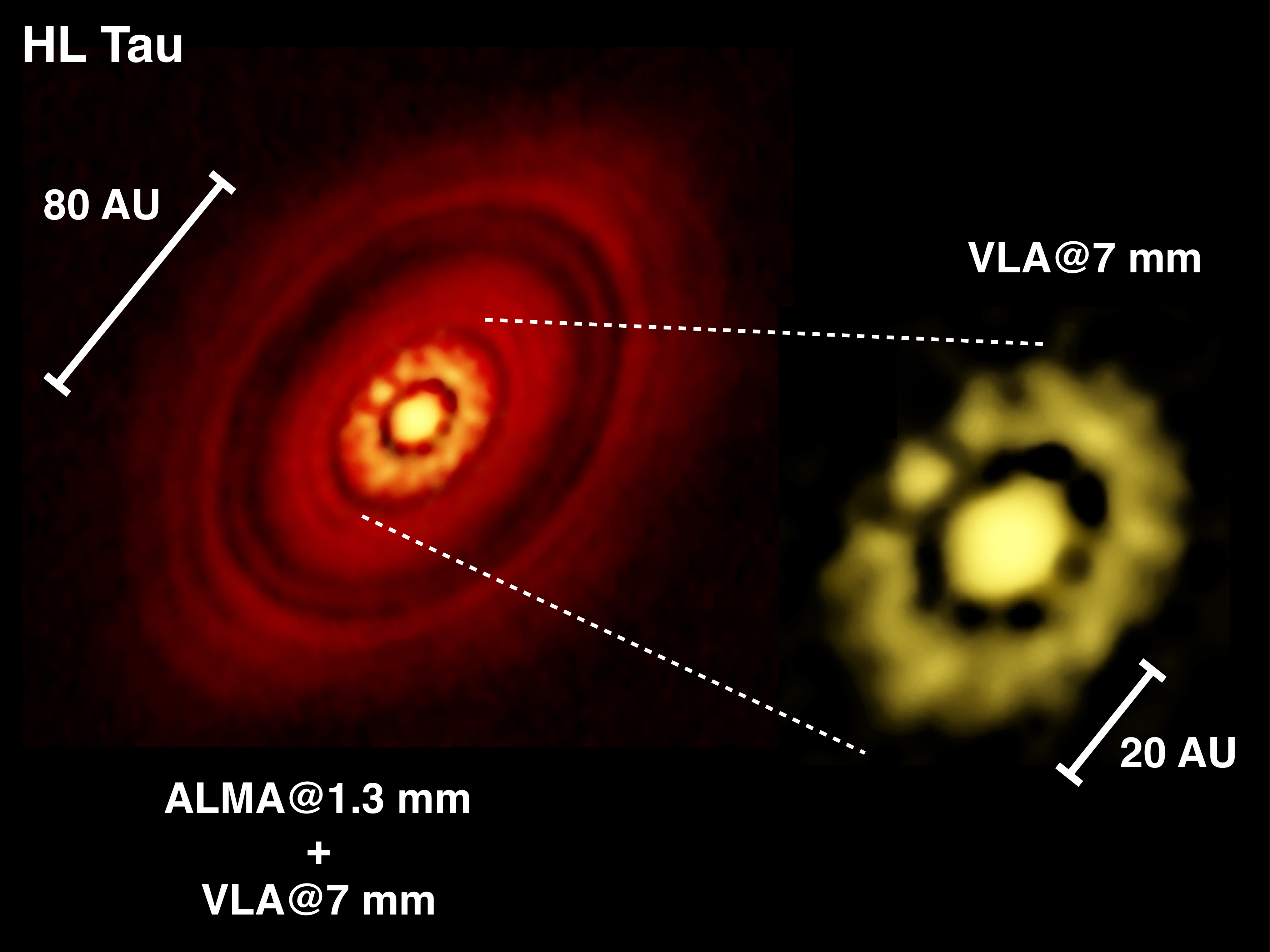}
\end{center}
\caption{A comparison of the ALMA 230\,GHz ({\it left}; \citealt{brogan15}) and VLA 43\,GHz ({\it right}; \citealt{carrasco-gonzalez16}) continuum images for the HL Tau disk.  At the lower frequency of the VLA data, the emission preferentially tracers structures in the inner disk but the observations are very expensive and only possible for the very brightest targets (and are still limited in resolution).  The ngVLA will make substantially improved measurements for a large population of disks routine.  ({\it credit: C.~Carrasco-Gonzalez / Bill Saxton / NRAO / AUI / NSF})}
\end{figure*}

The solution to these potential problems is to instead probe these continuum substructures at lower frequencies ($\lesssim$100\,GHz), where the optical depths are correspondingly lower (a linear frequency scaling for the optical depth, $\tau_\nu \propto \nu$, is reasonable; e.g., see \citealt{ricci10}).  The current VLA is too limited in sensitivity and angular resolution to optimally perform in this area, but nevertheless has produced some early results that demonstrate feasibility for the brightest available targets \citep[e.g.,][]{isella14,marino15,macias16,macias17}.  Figure 3 illustrates perhaps the best example, for the HL Tau disk \citep{carrasco-gonzalez16}.  Perhaps the most compelling result from these forays are the hints of substructure asymmetries that are not visible at higher frequencies due to their high optical depths.  

Real progress in this line of inquiry into planetesimal formation will require the sensitivity and resolution capabilities only available with the proposed ngVLA facility.  The ideal tracer of the solids concentrated in protoplanetary disk substructures is the 30--100\,GHz continuum, which strikes the best balance in sensitivity (emission still bright), optical depth (low enough to reliably estimate densities), and angular resolution (high enough to resolve fine-scale features at disk radii as small as $\sim$1\,au).  A modest survey could be used to understand the underlying physical mechanisms that control the prevalence, forms, scales, amplitudes, spacings, and symmetry of disk substructures and their presumably crucial roles in the planet formation process.

\section{Benchmarks for Measuring Disk Substructures}

The experiment design principle most relevant for probing disk substructures is ultimately related to the feasibility of detecting a microwave continuum brightness temperature contrast for a given perturbation in the local surface density of solids (or, more precisely, the associated optical depth).  There is a bewildering array of potential {\it forms} for such perturbations -- ranging from simple rings and gaps, to spirals, to azimuthally asymmetric features like vortices, and perhaps even stochastically-distributed pockets of material -- depending on the various physical mechanism(s) responsible for modifying the gas pressure gradient.  Aside from theoretical ideas and the preliminary work being done with ALMA, we simply do not yet know the diversity of these forms in the general disk population.  So, to simplify a discussion of the tractability of studying these features, specifying the morphology is less important than the locations, spatial scales, and amplitudes of the associated particle concentrations.   

The baseline temperatures $T_d$ and optical depths $\tau_\nu$, and thereby brightness temperatures $T_b(\nu)$, should typically be higher in the inner disk (closer to the host star).  As some fiducial reference points to illustrate feasibility, we consider hypothetical substructures at radii of 5 and 40\,au in a disk model orbiting a $\sim$1\,$M_\odot$ host star at an age of 1\,Myr (with a corresponding $L_\ast \approx 2$\,$L_\odot$).  We assume that the baseline 230\,GHz continuum optical depths are 10 and 1 at these locations, based roughly on models of the HL Tau and TW Hya ALMA studies referenced in the previous section, and that the optical depth spectra vary like $\tau_\nu \propto \nu^\beta$ with $\beta \approx 1$.  Given a simple model of the stellar irradiation, we expect temperatures of $T_d \approx 75$ and 25\,K at 5 and 40\,au, respectively.  These basic properties also set the expected scale of local perturbations, thought to be tied to the gas pressure scale height $H_p = c_s / \Omega$, the ratio of the sound speed to the Keplerian orbital frequency.  At 5 and 40\,au in this reference model, $H_p \approx 0.3$ and 2.5\,au (i.e., the aspect ratio is $H_p / r \approx 0.06$).  Presuming a Gaussian perturbation and a typical distance for nearby protoplanetary disks of 140\,pc, we will consider substructures with FWHM sizes of 3$H_p$ at 5\,au ($\sim$10\,mas), and $\sim$$H_p$ at 40\,au ($\sim$40\,mas), respectively.

\begin{figure*}[!t]
\includegraphics[width=\linewidth]{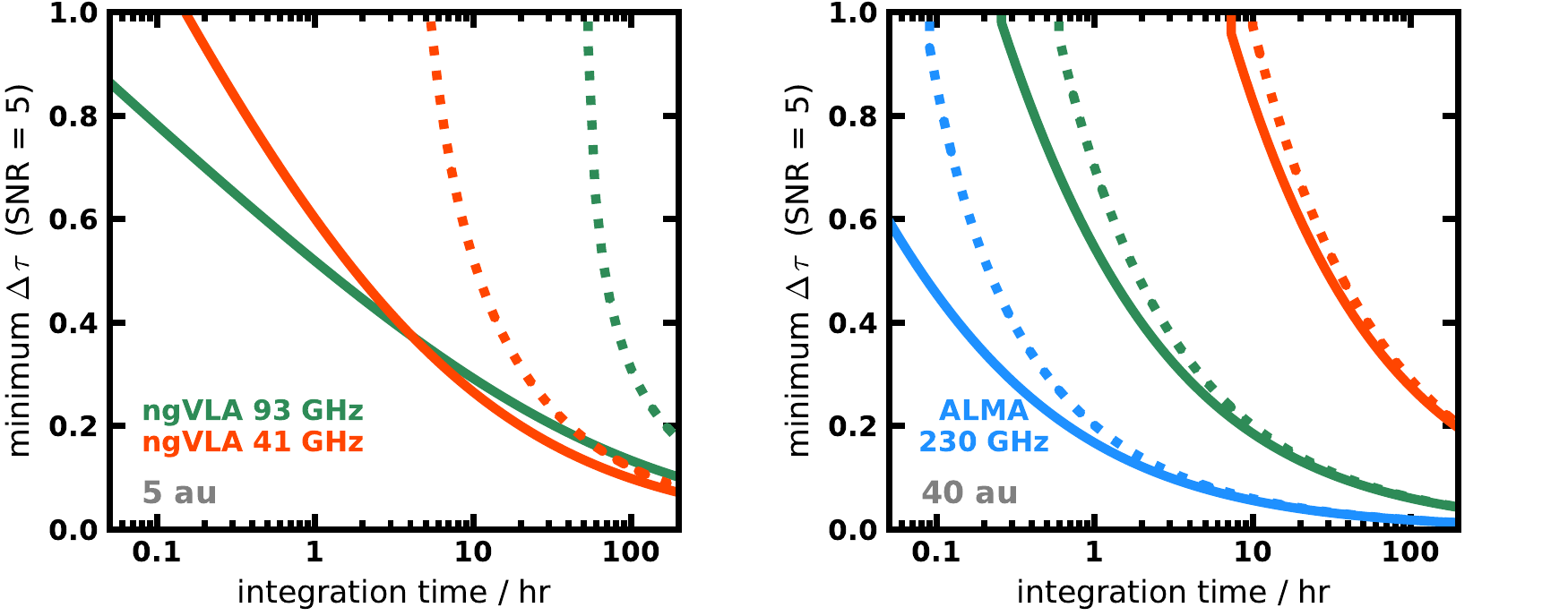}
\caption{Illustrative examples of the minimum optical depth (fractional) contrast that can be detected at 5\,$\sigma$ in a given integration time, for fiducial disk substructures at 5 ({\it left}) and 40\,au ({\it right}).  Solid curves are for $\tau_\nu$ depletions; dotted curves are for enhancements.  The blue, green, and red curves are for ALMA 230\,GHz, ngVLA 93\,GHz, and ngVLA 41\,GHz, respectively.  See the text for a detailed discussion.}
\end{figure*}

With these assumptions and some measured or expected sensitivity and resolution metrics, we can then calculate the minimum detectable (at a given signal-to-noise ratio, SNR) fractional optical depth perturbation for a given integration time ($t_{\rm int}$).  Presuming the local particle opacity does not also change (which is not necessarily the case), there is a one-to-one correspondance between optical depth and surface density perturbations.  Because the base optical depths at these locations and frequencies span around unity, there is some nuance in the meaning of `fractional optical depth perturbation' depending on whether we are considering an enhancement, $\tau_\nu^\prime = \tau_\nu (1 + \Delta \tau_\nu)$, or a depletion, $\tau_\nu^\prime = \tau_\nu (1 - \Delta \tau_\nu)$: we will illustrate both scenarios for clarity.  However, in either case, we can quantify the sensitivity by first calculating a generic
\begin{equation}
    {\rm SNR}_\nu = \frac{1}{\sigma_\nu} \, \left(\frac{t_{\rm int}}{2}\right)^{1/2} \, 
    \left| \, T_d \, (1 - e^{-\tau_\nu}) - T_d \, (1 - e^{-\tau_\nu^\prime}) \, \right| ,
\end{equation}
where $\sigma_\nu$ is the RMS noise level for a given beam area (in $T_b$ units), as a function of $\Delta \tau_\nu$ and $t$.  From that, we can then identify the smallest fractional optical depth contrast (i.e., the minimum $\Delta \tau_\nu$) that could be detected with a specific SNR threshold for a given integration time $t_{\rm int}$.  Considering ALMA observations at 230\,GHz (Band 6) and ngVLA measurements at 93 (W-band) and 41\,GHz (Q/Ka-band), we collect estimates of $\sigma_\nu$ for FWHM beam diameters of 10 and 40\,mas from the ALMA sensitivity calculator\footnote{\url{https://almascience.eso.org/proposing/sensitivity-calculator}} and the metrics discussion in ngVLA Memo 17\footnote{\url{http://library.nrao.edu/public/memos/ngvla/NGVLA_17.pdf}}.  Figure 3 shows the 5\,$\sigma$ (SNR = 5) optical depth ``contrast curves" for the 5 and 40\,au reference substructures.  The solid and dotted curves show the cases of optical depth depletion and enhancement, respectively.

These contrast curves in Figure 3 are only illustrative examples, but they demonstrate the general feasibility of ngVLA observations that should play substantial roles in the study of protoplanetary disk substructures.  As an approximate point of reference, gas density (thereby pressure) perturbations of $\sim$20\%\ induced by interactions with a $\sim$0.1\,$M_{\rm jup}$ planet \citep{fung14}, MHD zonal flows \citep{simon14}, or weak vortices \citep{goodman87} are expected to be sufficient to trap particles in a nominal disk.  Moreover, the corresponding perturbations in the local surface density of disk solids could be considerably amplified relative to the modulations in the gas densities \citep[e.g.,][]{paardekooper06,pinilla12a}.  

At intermediate disk radii ($\sim$tens of au; e.g., the right panel of Fig.~3), where optical depths and temperatures are low but substructures should be relatively larger, there are opportunities for synergy with high frequency measurements from ALMA at matching angular resolution.  For substructures at these distances from the host star, even relatively modest optical depth contrasts could be detected and resolved in a reasonable ngVLA observing time down to frequencies of $\sim$30\,GHz.  For optically thicker (or warmer) cases, it would be possible to push down to 10\,GHz, and thereby more directly probe concentrations of the largest accesible particle sizes ($\sim$few cm).  

The real benefit of the ngVLA is more apparent at small disk radii (inside $\sim$10\,au; e.g., the left panel of Fig.~3), where optical depths and temperatures are higher and substructures should be smaller.  There, the combination of long baselines and low frequencies offered by the ngVLA provide {\it unique} access to inner disk substructures on the smallest spatial scales.  For example, $\sim$20\%\ density depletions at 5\,au are accessible at $\sim$30--100\,GHz in $\sim$10--30\,hours of ngVLA integration, suitable for detecting the gaps in the local distribution of solids that are dynamically carved by young, super-Earth  planets \citep[e.g.,][]{ricci18}.  At the higher frequencies probed by ALMA, even relatively large density contrasts could remain hidden for even large integration times if the background optical depths are too high.  Moreover, beam dilution due to limited baseline lengths creates an additional ALMA barrier; in short, there is good reason that no 230\,GHz contrast curve is shown in the left panel of Figure 3.

\section{Quantifying Substructures}

The detectability of disk substructures is obviously fundamental, but it is not the sole value of ngVLA in this scientific context.  Resolved wideband continuum measurements in the 30--100\,GHz range permit more robust estimates of small-scale spatial variations in the continuum optical depth and spectral morphology (e.g., within and between the $\sim$20\,GHz-wide high-frequency ngVLA receiver bands), which in turn can provide strong constraints on the amplitude and shape of the particle size distribution and density perturbation of individual substructures.  The spatial morphology of the spectrum across such a feature could reveal new insights on the hydrodynamical nature of particle traps: for example, various vortex models make different predictions about where and how strongly particles of different sizes are concentrated with respect to the local gas pressure maximum \citep[e.g.,][]{baruteau16,sierra17}.  The spectral dependence of the brightness temperature perturbation from a substructure is itself sensitive to the underlying particle density contrast: if high concentrations of ``pebbles" can be inferred, we could rule on the likelihood that fast-acting mechanisms (e.g., the streaming instability; \citealt{youdin05}) are indeed the dominant pathways for the formation of planetesimals.  In all of these aspects, the ngVLA will be an indispensable and ground-breaking tool for these lines of work. 

Finally, the unique ability of the ngVLA to probe substructure asymmetries at small disk radii creates some new opportunities in the time-domain.  Long-term monitoring observations would be of extraordinarily high value for any inner disk substructures with a clear azimuthal asymmetry.  The motions of such features compared to their expected Keplerian orbital rates can help illuminate the natures of the underlying source of the pressure maximum responsible for the particle trap.  As a reference point, a Keplerian orbital displacement of a full 10\,mas beam would take $\sim$6 months for a structure at a radius of 5\,au in a disk orbiting a solar-mass host.

\section{Conclusions}

The characterization of small-scale substructures in the spatial distributions of protoplanetary disk solids is the most pressing issue at the forefront of observational planet formation research.  The new capabilities in high resolution wideband radio continuum imaging promised by the ngVLA will enable rapid progress in the field, delivering new insights on the particle size distributions in local concentrations of solids and offering unique access to the smallest substructure features at the locations where terrestrial planets (and most giant planets) are expected to form.  In doing so, the ngVLA measurements have some natural synergy with ALMA data at higher frequencies, facilitating more robust constraints on particle size distributions and better links with resolved tracers of the molecular gas.  Moreover, ngVLA observations of inner disk substructures will prove to be an important complement for tracers of inner disk material probed with the {\it JWST}, offering a structural template upon which to interpret those spatially unresolved infrared continuum and molecular spectra data.  Overall, the ngVLA should prove to be a crucial tool for better understanding the complex metamorphosis of circumstellar disks into young planetary systems.

%\acknowledgements ...  % Keep this text on the same line as the \verb"\acknowledgements" command because it makes things a lot easier.


\begin{thebibliography}{}

\bibitem[ALMA Partnership et al.(2015)]{brogan15} ALMA Partnership, et al. 2015, \apjl, 808, L3

\bibitem[Andrews et al.(2016)]{andrews16} Andrews, S.~M., et al. 2016, \apjl, 820, L40

\bibitem[Baruteau et al.(2014)]{baruteau14} Baruteau, C., et al. 2014, in Protostars and Planets VI, eds.~H. Beuther, R.~S.~Klessen, C.~P.~Dullemond, \& Th.~Henning (Univ.~Arizona Press: Tucson), 667

\bibitem[Baruteau \& Zhu(2016)]{baruteau16} Baruteau, C., \& Zhu, Z. 2016, \mnras, 458, 3927

\bibitem[Brauer et al.(2007)]{brauer07} Brauer, F., Dullemond, C.~P., Johansen, A., Henning, Th., Klahr, H., \& Natta, A. 2007, \aap, 469, 1169

\bibitem[Carrasco-Gonz{\'a}lez et al.(2016)]{carrasco-gonzalez16} Carrasco-Gonz{\'a}lez, C., et al. 2016, \apjl, 821, L16

\bibitem[Chiang \& Youdin(2010)]{chiang10} Chiang, E., \& Youdin, A.~N. 2010, AREPS, 38, 493

\bibitem[Cieza et al.(2016)]{cieza16} Cieza, L.~A., et al. 2016, Nature, 535, 258

\bibitem[Cieza et al.(2017)]{cieza17} Cieza, L.~A., et al. 2017, \apjl, 851, L23

\bibitem[Cox et al.(2017)]{cox17} Cox., E.~G., et al. 2017, \apj, 851, 83

\bibitem[Dipierro et al.(2018)]{dipierro18} Dipierro, G., et al. 2018, \mnras, 475, 5296

\bibitem[Dzyurkevich et al.(2010)]{dzyurkevich10} Dzyurkevich, N., Flock, M., Turner, N., Klahr, H., \& Henning, Th. 2010, \aap, 515, 70

\bibitem[Fedele et al.(2017)]{fedele17} Fedele, D., et al. 2017, \aap, 600, 72

\bibitem[Fedele et al.(2018)]{fedele18} Fedele, D., et al. 2018, \aap, 610, 24

\bibitem[Flock et al.(2015)]{flock15} Flock, M., Ruge, J.~P., Dzyurkevich, N., Henning, Th., Klahr, H., \& Wolf, S. 2015, \aap, 574, 68

\bibitem[Fung et al.(2014)]{fung14} Fung, J., Shi, J.-M., \& Chiang, E. 2014, \apj, 782, 88

\bibitem[Goodman et al.(1987)]{goodman87} Goodman, J., Narayan, R., \& Goldreich, P. 1987, \mnras, 225, 695

\bibitem[Huang et al.(2018)]{huang18} Huang, J., et al. 2018, \apj, 852, 122

\bibitem[Isella et al.(2014)]{isella14} Isella, A., Chandler, C.~J., Carpenter, J.~M., P{\'e}rez, L.~M., \& Ricci, L. 2014, \apj, 788, 129

\bibitem[Isella et al.(2016)]{isella16} Isella, A., et al. 2016, Phys.~Rev.~Lett., 117, 251101

\bibitem[Johansen et al.(2014)]{johansen14} Johansen, A., Blum, J., Tanaka, H., Ormel, C.~W., Bizarro, M., \& Rickman, H. 2014, in Protostars and Planets VI, eds.~H. Beuther, R.~S.~Klessen, C.~P.~Dullemond, \& Th.~Henning (Univ.~Arizona Press: Tucson), 547

\bibitem[Loomis et al.(2017)]{loomis17} Loomis, R.~A., {\"O}berg, K.~I., Andrews, S.~M., \& MacGregor, M.~A. 2017, \apj, 840, 23

\bibitem[Mac{\'\i}as et al.(2016)]{macias16} Mac{\'\i}as, E., et al. 2016, \apj, 829, 1

\bibitem[Mac{\'\i}as et al.(2017)]{macias17} Mac{\'\i}as, E., et al. 2017, \apj, 838, 97

\bibitem[Marino et al.(2015)]{marino15} Marino, S., et al. 2015, \apj, 813, 76

\bibitem[Paardekooper \& Mellema(2006)]{paardekooper06} Paardekooper, S.-J., \& Mellema, G. 2006, \aap, 453, 1159

\bibitem[P{\'e}rez et al.(2016)]{perez16} P{\'e}rez, L.~M., et al. 2016, Science, 353, 1519

\bibitem[Pinilla et al.(2012)]{pinilla12a} Pinilla, P., Birnstiel, T., Ricci, L., Dullemond, C.~P., Uribe, A.~L., Testi, L., \& Natta, A. 2012, 538, 114

\bibitem[Pollack et al.(1996)]{pollack96} Pollack, J.~B., Hubickyj, O., Bodenheimer, P., Lissauer, J.~J., Podolak, M., \& Greenzweig, Y. 1996, Icarus, 124, 62

\bibitem[Raymond et al.(2014)]{raymond14} Raymond, S.~N., Kokubo, E., Morbidelli, A., Morishima, R., \& Walsh, K.~J. 2014, in Protostars and Planets VI, eds.~H. Beuther, R.~S.~Klessen, C.~P.~Dullemond, \& Th.~Henning (Univ.~Arizona Press: Tucson), 595

\bibitem[Ricci et al.(2010)]{ricci10} Ricci, L., Testi, L., Natta, A., Neri, R., Cabrit, S. \& Herczeg, G.~J. 2010, \aap, 512, 15

\bibitem[Ricci et al.(2018)]{ricci18} Ricci, L., Liu, S.-F., Isella, A., \& Li, H. 2018, \apj, 853, 110

\bibitem[Sierra et al.(2017)]{sierra17} Sierra, A., Lizano, S., \& Barge, P. 2017, \apj, 850, 115

\bibitem[Simon \& Armitage(2014)]{simon14} Simon, J.~B., \& Armitage, P.~J. 2014, \apj, 784, 15

\bibitem[Stammler et al.(2017)]{stammler17} Stammler, S.~M., Birnstiel, T., Pani{\'c}, O., Dullemond, C.~P., \& Dominik, C. 2017, \aap, 600, 140

\bibitem[Takeuchi \& Lin(2002)]{takeuchi02} Takeuchi, T., \& Lin, D.~N.~C. 2002, \apj, 581, 1344

\bibitem[Tsukagoshi et al.(2016)]{tsukagoshi16} Tsukagoshi, T., et al. 2016, \apjl, 829, L35

\bibitem[van der Plas et al.(2017)]{vanderplas17} van der Plas, G., et al. 2017, \aap, 597, 32

\bibitem[Whipple(1972)]{whipple72} Whipple, F.~L. 1972, in From Plasma to Planet, ed.~A.~Elvius, 211

\bibitem[Youdin \& Goodman(2005)]{youdin05} Youdin, A.~N., \& Goodman, J. 2005, \apj, 620, 459

\bibitem[Zhang et al.(2016)]{zhang16} Zhang, K., et al. 2016, \apjl, 818, L16

\bibitem[Zhu et al.(2014)]{zhu14} Zhu, Z., Stone, J.~M., Rafikov, R.~R., \& Bai, X. 2014, \apj, 785, 122

\end{thebibliography}
\end{document}